\documentclass{article}
\usepackage{amssymb}
\def\tr{{\rm tr}}
\def\R{{\mathbb R}}
\def\Z{{\mathbb Z}}
\def\C{{\mathbb C}}
\def\P{{\mathbb P}}
\newtheorem{theorem}{Theorem}

\begin{document}

\title{Future geodesic completeness of some spatially homogeneous
solutions of the vacuum Einstein equations in higher dimensions} 

\author{Arne G\"odeke and Alan D. Rendall \\ 
Max-Planck-Institut f\"ur Gravitationsphysik\\
Albert-Einstein-Institut\\
Am M\"uhlenberg 1\\
14476 Potsdam, Germany}

\date{}

\maketitle

\begin{abstract}
It is known that all spatially homogeneous solutions of the vacuum Einstein
equations in four dimensions which exist for an infinite proper time 
towards the future are future geodesically complete. This paper 
investigates whether the analogous statement holds in higher dimensions. A
positive answer to this question is obtained for a large class of models
which can be studied with the help of Kaluza-Klein reduction to solutions
of the Einstein-scalar field equations in four dimensions. The proof of this
result makes use of a criterion for geodesic completeness which is applicable
to more general spatially homogeneous models.
\end{abstract}

\section{Introduction}\label{intro}

The solutions of the Einstein equations most commonly applied in cosmology
are spatially homogeneous and include some that recollapse and some that 
expand for ever. In the former case the proper length of the timelike curves
orthogonal to the hypersurfaces of homogeneity (the worldlines of comoving
observers) is finite. Along a curve of this type the expansion of this 
congruence of curves is first positive, then zero and finally negative. This 
is what is meant by recollapse. In the latter case the expansion is always 
positive and the length of the curves orthogonal to the hypersurfaces of 
homogeneity is infinite towards the future. This is what is meant by saying 
that the solution expands for ever.

On physical grounds it is reasonable to expect that a forever expanding 
solution will be future geodesically complete, i.e. there should be no 
singularities in the future. A theorem which confirms this expectation
is proved in \cite{rendall95a} (Theorem 2.1 of that paper). In order to 
explain this result let $t$ be a Gaussian time coordinate based on one of the 
hypersurfaces of homogeneity and let $\tr k$ be the mean curvature of these 
hypersurfaces. Then $\tr k$ is a function of $t$. The expansion of the 
congruence of timelike geodesics mentioned above is $-\frac13\tr k$.
Define an expanding phase of the solution to be a time interval $(t_1,t_2)$
on which $\tr k<0$. Here $t_1$ and $t_2$ may be finite or infinite. The
theorem says that, under certain assumptions, if $t_2<\infty$ the solution
can be extended to a time interval $(t_1,t_3)$ with $t_3>t_2$. Moreover, if
$t_2=\infty$ then the spacetime is future geodesically complete. The theorem
is formulated for spatially homogeneous solutions of the Einstein equations 
coupled to a matter model satisfying three assumptions. Two of these 
are inequalities on the energy-momentum tensor, the dominant energy and 
non-negative sum pressures conditions. The third is a continuation property 
for solutions of the Einstein-matter equations. The only thing about these 
conditions which is important in what follows is that they are all 
satisfied in the case of the Einstein vacuum equations. Hence the theorem 
applies in the vacuum case.

In a spatially homogeneous spacetime, by definition the isometry group acts
transitively on the hypersurfaces of homogeneity. In four dimensions there
is a well-known classification of the possible isometry groups. Either the
group can be taken to have dimension three or a higher-dimensional group
is necessary. In the second of these cases there is only one possibility,
which is called Kantowski-Sachs symmetry. In the first case any connected 
three-dimensional Lie group can occur. For the purposes of what follows it
may be assumed without loss of generality that the spacetime is simply
connected. This is because passing to the universal cover of a given spacetime 
has no effect on the dynamics. With this assumption the hypersurfaces of 
homogeneity can be identified with the group itself and the induced metric on 
one of these hypersurfaces with a left-invariant metric on that group. The 
classification of connected and simply connected Lie groups reduces to that 
of the corresponding Lie algebras. The three-dimensional Lie algebras were 
classified by Bianchi into types I to IX. It turns out that a maximally 
extended vacuum spacetime of Bianchi type I-VIII expanding at some time is 
forever expanding while one which is of type IX or has Kantowski-Sachs 
symmetry recollapses \cite{linwald}. The 
cases where recollapse takes place are precisely those where there is a 
spatial metric in the class whose scalar curvature $R$ is positive. That this 
is a necessary condition for recollapse is not hard to see. The Hamiltonian 
constraint on a hypersurface of homogeneity is (in the vacuum case)
\begin{equation}
R-k_{ab}k^{ab}+(\tr k)^2=0
\end{equation} 
where $k_{ab}$ is the second fundamental form. Recollapse means that there
is a maximal hypersurface, i.e. one where $\tr k=0$. There $R$ is clearly
non-negative. If it is zero then $k_{ab}=0$. Moreover it can be shown that 
if $R=0$ the induced metric of that hypersurface is flat and that the 
solution arising from those data is a flat spacetime where $\tr k$ vanishes 
at all times. The latter case cannot occur in the context of spatially 
homogeneous cosmological models which are expanding at some time. 

The aim of this paper is to investigate to what extent the results just 
stated for four-dimensional spatially homogeneous vacuum spacetimes extend 
to higher dimensions $n+1$. In particular, sufficient conditions for future 
geodesic completeness are derived. Only the case where the isometry group 
can be taken to be of dimension $n$ is considered. This is the
analogue in higher dimensions of Bianchi models. The analogue of 
Kantowski-Sachs models is not considered. The same argument as in four 
dimensions applies to show that a necessary condition for recollapse is the 
existence of a left-invariant metric of positive scalar curvature on the
group of interest. This motivates us to restrict attention to Lie groups
admitting no left-invariant metric of positive scalar curvature. In this
case it has been shown that vacuum spacetimes with this type of symmetry
exist for infinite Gaussian time in the future (\cite{rendall08}, Theorem
5.3). On the other hand it turns out that the arguments used to prove
geodesic completeness in the four-dimensional case do not easily generalize
to higher dimensions. In what follows partial results on this question are
obtained.

For a general dimension there is no known analogue of the Bianchi 
classification of Lie algebras. In some of the lower dimensions there are
results. A review of what is known in the case of four-dimensional Lie
algebras can be found in \cite{hervik02}. The strategy in the following is
to identify some classes of Lie groups for which geodesic completeness can
be proved. The main case considered is that where there is an isometry group
of the form $G_3\times (\R\rtimes \Z_2)^{n-3}$ for a three-dimensional 
group $G_3$. The spatial manifold is given by the identity component of this 
group, $G_3\times \R^{n-3}$. The strategy for handling these groups is to use 
Kaluza-Klein reduction. The presence of the discrete symmetry defined by
the group $\Z_2^{n-3}$ ensures that the result is the Einstein equations 
coupled to linear scalar fields in four dimensions. Without it a more 
complicated reduced system would be obtained. In the case $n=4$, for instance, 
it would be the Einstein-Maxwell-dilaton system. Section 2 contains the basic 
notation and equations, including those for the Kaluza-Klein reduction. In 
the third section results on the Einstein-scalar field system in four 
dimensions are proved. These are used in treating higher-dimensional vacuum
spacetimes in later sections and are also of interest in their own right.
A general criterion for geodesic completeness is stated and 
proved in Section 4. It is then combined with the results on the scalar
field to obtain the main theorem, Theorem \ref{maintheorem}. Section 5 
discusses geodesic completeness in some examples with other types of 
symmetry and illustrates how the results of Section 4 can be applied.
Conclusions and an outlook are contained in Section 6.

This paper is based in part on the diploma thesis of the first author 
\cite{goedeke}.

\section{The basic equations}

Let $G$ be a connected and simply connected Lie group and $\theta^a$ a
basis of the space of left-invariant one-forms on $G$. Consider the metric
\begin{equation}
-dt^2+g_{ab}(t)\theta^a\otimes\theta^b
\end{equation}
on the manifold $[t_0,\infty)\times G$ for some constant $t_0$. Let 
$k_{ab}$ be the components in the given basis of the second fundamental form 
of the level hypersurfaces of $t$. By the definition of the second 
fundamental form 
\begin{equation}
\partial_t g_{ab}=-2k_{ab}.
\end{equation} 
It will now be supposed that this metric satisfies the Einstein vacuum
equations and these will be decomposed using the foliation by the level
hypersurfaces of $t$. The Einstein constraint equations are
\begin{eqnarray}
&&R-k_{ab}k^{ab}+(\tr k)^2=0,    \\
&&\nabla_a k^a{}_b=0.
\end{eqnarray}
The evolution equations are given by
\begin{equation}
\partial_t k^a{}_b=(\tr k)k^a{}_b+R^a{}_b.
\end{equation}
A real number $\lambda$ is said to be an eigenvalue of $k_{ab}$ with respect 
to $g_{ab}$ if there is a non-zero vector $v^a$ which satisfies the equation
$k_{ab}v^b=\lambda g_{ab}v^b$. In $n$ space dimensions there are $n$ 
eigenvalues of $k_{ab}$ with respect to $g_{ab}$, counting multiplicity.
Call them $\lambda_a$. The mean curvature $\tr k=\sum_a\lambda_a $ never
vanishes in an expanding phase. Define the generalized Kasner exponents (GKE)
by $p_a=\frac{\lambda_a}{\tr k}$. When the Lie group $G$ is Abelian the $p_a$
are independent of time and the solution can be written explicitly as
\begin{equation}
-dt^2+\sum_{a=1}^n t^{2p_a}(dx^a)^2
\end{equation}
where $\sum_{a=1}^n p_a=1$ and $\sum_{a=1}^n p_a^2=1$. In the case $n=3$ this 
is the Kasner solution.

These equations will now be specialized to the case that the Lie group is
of the form $G_3\times \R^{n-3}$ for a three-dimensional group $G_3$ and the
left-invariant basis $\theta^a$ is adapted to the product decomposition.
Call the basis vectors tangent to the first factor $\theta^i$, $i=1,2,3$.
The basis vectors corresponding to the second factor commute and so can be
chosen to be of the form $dy^I$, $I=1,\ldots,n-3$. Suppose further that
the action of the group extends to an action of 
$G_3\times (\R\rtimes \Z_2)^{n-3}$ which preserves the hypersurfaces of
constant $t$ and the product decomposition of the initial hypersurface. 
Then the generators of $\Z_2^{n-3}$ act by commuting reflections of
$\R^{n-3}$. Without loss of generality they may be assumed to be given 
by reflections in the coordinates $y^I$. The invariance of the metric
under these transformations implies that it can be written in the form
\begin{equation}
-dt^2+g_{ij}(t)\theta^i\otimes\theta^j+\sum_ie^{2\phi^I(t)}(dy^I)^2
\end{equation} 
for scalar functions $\phi^I$. Let $\phi=\sum_I\phi^I$ and write 
$g_{ij}=e^{-\phi}\tilde g_{ij}$. Let $\tilde t$ be a Gaussian coordinate for 
the conformally rescaled metric. Then $\frac{d\tilde t}{dt}=e^{\frac{\phi}{2}}$.
Rewriting the Einstein equations in terms of $\tilde g_{ij}$ and 
$\phi^I$ gives the equations
\begin{eqnarray}
&&-\tilde k_{ij}\tilde k^{ij}+(\tr \tilde k)^2=\rho,         \\
&&\partial_{\tilde t} \tilde k^i{}_j=(\tr \tilde k)\tilde k^i{}_j
+\tilde R^i{}_j-\rho\delta^i_j, \\
&&\partial_{\tilde t}^2\phi^I-(\tr \tilde k)\partial_{\tilde t}\phi^I=0.
\end{eqnarray} 
Here
\begin{equation}
\rho=\frac12\sum_I (\partial_{\tilde t}\phi^I)^2.
\end{equation}
These equations are identical to the Einstein equations in four dimensions
coupled to $n-3$ non-interacting massless scalar fields. For prescribed 
initial data it is possible to do a rotation in $\R^{n-3}$ so as to set
all $\partial_{\tilde t}\phi^I$ except $\partial_{\tilde t}\phi^1$ to zero. By 
the equations of motion for 
the scalar fields all $\phi^I$ except $\phi^1$ will be constant during the 
evolution. Then only $\phi^1$ makes a contribution to the energy-momentum
tensor. The essential dynamics is that of a single scalar field. The
four-dimensional metric and the scalar field $\phi^1$ define a 
five-dimensional metric. The full spacetime is obtained by taking the 
product of the five-dimensional metric with a flat time-independent metric.

The generalized Kasner exponents of the $(n+1)$-dimensional metric are
related to the four-dimensional quantities by
\begin{eqnarray}
&&p_a=\frac{\tilde\lambda_a+\frac12e^{-\frac{\phi}{2}}\partial_t\phi}
{\tr\tilde k+\frac12e^{-\frac{\phi}{2}}\partial_t\phi},\label{red1}    \\
&&p_I=\frac{-e^{-\frac{\phi}{2}}\partial_t\phi^I}{\tr \tilde k
+\frac12e^{-\frac{\phi}{2}}\partial_t\phi}.
\label{red2}
\end{eqnarray}

In a spatially homogeneous situation a scalar field is equivalent to an
untilted stiff fluid. Thus to understand the dynamics of the scalar field
case known results for a stiff fluid can be applied. The matter variables
of the two models are related by $\rho=\frac12(\partial_{\tilde t}\phi)^2$.
Note that $e^{-\frac{\phi}{2}}\partial_t=\partial_{\tilde t}$.

It was pointed out in the introduction that it is important whether
a given Lie group $G$ admits left-invariant metrics with positive scalar 
curvature. If $G$ is a Lie group of any of the Bianchi types except
IX it admits no metric of this kind. This is equivalent, for simply 
connected Lie groups, to the property that they are diffeomorphic to
Euclidean space \cite{berardbergery}. It follows that a Lie group with topology 
$G\times\R^{n-3}$ is also homeomorphic to Euclidean space and thus also 
admits no left-invariant metrics of positive scalar curvature. Hence groups 
of this type fail the necessary condition for recollapse mentioned in the 
introduction. This is a sign that they are good candidates for the symmetry 
groups of future geodesically complete metrics. 

\section{Four-dimensional models with a scalar field}

This section contains results on the late-time behaviour of solutions of 
the Einstein-scalar field system in four dimensions, some of which are 
deduced from results on untilted stiff fluids taken from the literature.
When information on the dynamics of the energy density is available 
the asymptotics of $\phi$ can be obtained by integration. Much of the analysis
is based on the Wainwright-Hsu formulation of the equations for Bianchi
models \cite{wainwright89}, which will now be recalled. Since the only case
of interest in the following is that of a stiff fluid the equations are only
written down for the case where the parameter $\gamma$ in the equation of
state $p=(\gamma-1)\rho$ of the fluid is equal to two. The system of 
evolution equations is
\begin{eqnarray}
&&N_1'=(q-4\Sigma_+)N_1,                         \\
&&N_2'=(q+2\Sigma_++2\sqrt{3}\Sigma_-)N_2,       \\
&&N_3'=(q+2\Sigma_+-2\sqrt{3}\Sigma_-)N_3,       \\
&&\Sigma_+'=-(2-q)\Sigma_+-3S_+,                 \\
&&\Sigma_-'=-(2-q)\Sigma_--3S_-,                 \\
&&\Omega'=-2(2-q)\Omega  
\end{eqnarray}
where
\begin{eqnarray}
&&q=2(\Omega+\Sigma_+^2+\Sigma_-^2),             \\
&&S_+=\frac12 [(N_3-N_2)^2-N_1(2N_1-N_2-N_3)],    \\
&&S_-=\frac12 (N_3-N_2)(N_1-N_2-N_3).
\end{eqnarray}
The prime denotes $\frac{d}{d\tau}$ and $\tau$ is related to the Gaussian
time $t$ by $\frac{d\tau}{dt}=-\frac13\tr k$. There is also the equation 
which expresses the Hamiltonian constraint in these variables which is 
\begin{equation}
\Omega+\Sigma_+^2+\Sigma_-^2+\frac34(N_1^2+N_2^2+N_3^2-2(N_1N_2+N_2N_3+N_1N_3))
=1.
\end{equation}
It will not be necessary to recall the definition of all the variables in
the system here. What is important is that $\Sigma_+$ and $\Sigma_-$ are
linear combinations of the generalized Kasner exponents which are linearly
independent. This means that if $\Sigma_+$ and $\Sigma_-$ tend to finite 
limits as $t\to\infty$ the same is true of $p_1$, $p_2$ and $p_3$. The density 
parameter $\Omega$ is equal to $\frac{3\rho}{(\tr k)^2}$. The mean curvature
can be recovered from a solution of the Wainwright-Hsu system using the 
equation
\begin{equation}
(\tr k)'=-(1+q)(\tr k).
\end{equation}

\begin{theorem}\label{mainscalar}
Let $(M,g)$ be a solution of the Einstein equations coupled to a scalar field
$\phi$ in four dimensions with symmetry of a Bianchi type other than IX or 
VI${}_{-\frac19}$. If the solution is expanding at some time and the time 
interval of definition of the solution is maximal then the upper limit of the 
interval is infinity and the generalized Kasner exponents and the density 
parameter $\Omega$ converge to limits as $t\to\infty$.
\end{theorem}

\noindent
{\bf Proof} 
Bianchi type I solutions are defined by the condition 
$N_1=N_2=N_3=0$. It follows from the Hamiltonian constraint that $q=2$ and 
that $\Sigma_+$, $\Sigma_-$ and $\Omega$ are constant. That the conclusion
of the theorem holds for Bianchi type II solutions, which are defined by the 
conditions $N_2=N_3=0$ and $N_1\ne 0$, was shown in Lemma 7.1 of 
\cite{ringstrom01a}. Bianchi type VI${}_0$ solutions are defined by the 
conditions $N_1=0$, $N_2>0$ and $N_3<0$. The evolution equation for 
$\Sigma_+$ reduces to 
\begin{equation}\label{sigmaplusvi0}
\Sigma_+'=-(2-q)(\Sigma_++1).
\end{equation}
It follows from the Hamiltonian constraint that $q<2$. This in turn implies
that $\Sigma_++1$ is strictly positive and that $\Sigma_+$ is strictly 
decreasing. Applying the monotonicity principle to $\Sigma_+$ shows that 
there can be no $\omega$-limit point with $q<2$. Since the solution stays 
in a compact region it follows that $q\to 2$ as $\tau\to\infty$ and that 
$N_1$ and $N_2$ tend to zero. Since $\Sigma_+$ and $\Omega$ are monotone and 
bounded both of them tend to limits as $\tau\to\infty$. By the Hamiltonian
constraint the same is true of $\Sigma_-$. Bianchi type VII${}_0$ solutions 
are defined by the conditions $N_1=0$, $N_2>0$ and $N_3>0$. Locally 
rotationally symmetric solutions of type VII${}_0$ are defined by the 
additional conditions  $N_2=N_3$ and $\Sigma_-=0$.  In this case $S_+$, 
$S_1$ are zero and $q=2$. Hence $\Omega$ and $\Sigma_+$ are constant. The 
proof for solutions of type VII${}_0$ which are not locally rotationally 
symmetric results from a small modification of the proof of Proposition 5 
of \cite{ringstrom01b}, which treats the corresponding question in the vacuum 
case. First it can be shown by an argument similar to that used for type 
VI${}_0$ case that $\Omega$ and $\Sigma_+$ are strictly monotone. This makes 
use of the equation
\begin{equation}
(N_2-N_3)'=(q+2\Sigma_+)(N_2-N_3)+2\sqrt{3}\Sigma_-(N_2+N_3)
\end{equation}
to treat the points where $\Sigma_+'$ vanishes. Since $\Omega$ and $\Sigma_+$ 
are also bounded they must converge as $t\to\infty$. It is not known a priori 
that the whole solution is bounded and so it may not have an $\omega$-limit 
point. Suppose that it does have such a point. Using the monotonicity
principle it can be shown that if $\tau_n$ is a sequence of times along which 
the solution converges then either $(\Sigma_-^2+(N_2-N_3)^2)(\tau_n)$ tends
to zero or $N_1+N_2$ does so. It can then be concluded as in 
\cite{ringstrom01b}, using the monotone function $Z_{-1}$ of that paper
that $\Sigma_-\to 0$ as $\tau\to\infty$. Note that $Z_{-1}$ is monotone in
the case with matter (see \cite{ringstrom01b}, Lemma 10.1). It remains to 
treat the case in which the solution has no $\omega$ limit point. Then
$N_1$ and $N_2$ must tend to infinity and it can be assumed without loss of
generality that the limit of $\Omega+\Sigma_+^2$ is less than one. In
particular this means that the limit of $\Sigma_+$ cannot be minus one.
Hence, by (\ref{sigmaplusvi0}) the quantity $2-q$ must be in $L^1$. It can, 
however, be shown by a slight modification of an argument of Ringstr\"om
(\cite{ringstrom01b}, proof of Proposition 5) that $2-q$ is not in $L^1$.  That 
argument uses two quantities called $x$ and $y$ and the necessary modification
is to replace them by
\begin{eqnarray}
&&x=\frac{\Sigma_-}{(1-\Sigma_+^2-\Omega)^{\frac12}},    \\
&&y=\frac{\sqrt{3}}{2}\frac{N_2-N_3}{(1-\Sigma_+^2-\Omega)^{\frac12}}.
\end{eqnarray}
Bianchi type VIII solutions are defined by the conditions $N_1<0$, $N_2>0$ 
and $N_3<0$. The result of the theorem in this case follows from Theorem
3.1 and Corollary 3.1 of \cite{horwood03}.

For models of Bianchi class B it follows from Proposition 5.3 of 
\cite{hewitt93} that the $\omega$-limit set of any solution lies in a set
$\bar{\cal L}_k$. The parameter $k$ distinguishes the different Bianchi 
types and is equal to $h^{-1}$ in types VI${}_h$ and VII${}_h$. For each 
value of $k$, the set $\bar{\cal L}_k$ consists of equilibrium solutions.
It is an open curve ${\cal L}_k$ together with its two endpoints. The 
$\omega$-limit set of a solution is in fact a single point of 
$\bar{\cal L}_k$. This can be seen as follows. A generic point of 
the curve is transversely hyperbolic and so if it belongs to the
$\omega$-limit set no other point can.  The remaining points form 
a discrete set and so by connectedness of the $\omega$-limit set it 
cannot contain more than one of them.  

\vskip 10pt\noindent
The fact that Bianchi type IX is excluded in the hypotheses of this theorem
is essential. For the metric product of a four-dimensional solution of 
Bianchi type IX with the real line fails to exist globally towards the 
future. On the other hand there is no reason to expect that type 
VI${}_{-\frac19}$ has to be excluded. It appears nevertheless that up to now 
no proof is available in that case. 

Some more precise information will now be given on the asymptotics of scalar
field models. Let ${\cal D}$ be the disk in the space with coordinates
$(\Sigma_+,\Sigma_-,N_1,N_2,N_3)$ defined by the conditions that the $N_i$ are 
all zero and $\Sigma_+^2+\Sigma_-^2<1$. Let ${\cal L}$ be the line defined by 
$N_1=0$, $N_2=N_3>0$, $\Sigma_+=-1$ and $\Sigma_-=0$. 

\begin{theorem}\label{finescalar}
Let $(M,g)$ be a solution of the Einstein equations coupled to a scalar field
in four dimensions with symmetry of Bianchi class A. If the Bianchi type is
I, II or VI${}_0$ then the variables in the Wainwright-Hsu system converge
to limits as $\tau\to\infty$. If the Bianchi type is VIII the $\omega$-limit 
set of each solution is empty. In type VII${}_0$ both types of behaviour 
occur. The sets of points which occur as $\omega$-limit points is the disk
${\cal D}$ in type I, the open subset of ${\cal D}$ defined by the inequality 
$\Sigma_+>\frac12$ in type II, the point of $\bar{\cal D}$ with 
$(\Sigma_+,\Sigma_-)=(-1,0)$ in type VI${}_0$ and $\cal L$ in LRS type
VII${}_0$. The limit of $\Omega$ is non-zero in types I and II and zero in
types VI${}_0$ and VIII.   
\end{theorem}

\noindent
{\bf Proof} 
The Bianchi type I solutions of the Wainwright-Hsu system with a stiff fluid
are time independent and are exactly the points of ${\cal D}$. It follows
from Theorem \ref{mainscalar} that any solution of type II converges to a 
point of $\cal D$ as $\tau\to\infty$. It is a consequence of the evolution 
equation for $N_1$ that this point must satisfy $\Sigma_+\ge 0$. A more
detailed analysis reveals that this inequality is strict. To see this 
note that all terms on the right hand side of the equations for Bianchi type 
II solutions have a common factor $N_1$. Omitting this factor leads to a
new system whose integral curves are the same as those of the original 
system in the region where $N_1\ne 0$. The new system is
\begin{eqnarray}
&&\dot N_1=2(1-2\Sigma_+)-\frac32 N_1^2,                 \\
&&\dot\Sigma_+=3\left(1-\frac12\Sigma_+\right)N_1,       \\
&&\dot\Sigma_-=-\frac32 N_1\Sigma_-.     
\end{eqnarray}
This system has a line of stationary solutions defined by the conditions
$N_1=0$ and $\Sigma_+=\frac12$. Linearizing the system about one of these
points reveals that it has the eigenvalues $0,\pm 3i$. Because of the 
reduction theorem (cf. \cite{rendall08}, section 5.6) the qualitative behaviour 
of the solutions can be determined by analysing the dynamics on a centre 
manifold. The system on the centre manifold is such that no solution can 
converge to the stationary point while staying in the region $N_1>0$
(cf. \cite{perko}, p. 144, Theorem 5) and this implies the desired result. 
Any point of ${\cal D}$ with $\Sigma_0>\frac12$ occurs as the $\omega$-limit 
point of a solution of type II since the linearization there has a negative 
eigenvalue. By Theorem \ref{mainscalar} any solution of type VI${}_0$ 
converges to a point of $\bar{\cal D}$. As a consequence of the evolution 
equations for $N_2$ and $N_3$ this can only be the point where $\Sigma_+=-1$. 
In particular $\Omega\to 0$ and the solution looks asymptotically like
a vacuum solution. Any locally rotationally symmetric solution of type 
VII${}_0$ has $\Sigma_-=0$, $N_2=N_3$ and a constant value of $\Sigma_+$.
As $\tau\to\infty$ the quantity $N_2$ tends to infinity. For type VII${}_0$ 
solutions which are not LRS it has already been shown that 
$\Sigma_+^2+\Omega\to 1$ and that $\Sigma_-$ and $N_2-N_3$ tend to zero.
If the solution has no $\omega$-limit point then $\Omega\to 0$ as 
$\tau\to\infty$ and $\Sigma_+\to 1$. For general type VIII it is shown in
\cite{horwood03} that $\Omega\to 0$ and $N_+\to\infty$.

\section{A criterion for geodesic completeness}

The aim of this section is to prove the following criterion for geodesic
completeness. The general set-up is a spatially homogeneous spacetime
defined by a one-parameter family $g_{ab}(t)$ of left-invariant metrics on
a Lie group $G$ of dimension $n$.

\begin{theorem}\label{gencomplete}
Let $(M,g)$ be a locally spatially homogeneous vacuum spacetime of dimension 
$n+1<10$ whose symmetry is defined by a Lie group $G$ which admits no
left-invariant metrics of positive scalar curvature. If for some $t_0$
the generalized Kasner exponents satisfy
\begin{equation}
\sup_{t\ge t_0}\min_a p_a(t)-\inf_{t\ge t_0}\min_a p_a(t)\le\frac{9-n}{4n}
\end{equation}
then the spacetime is geodesically complete.
\end{theorem} 

\noindent
{\bf Proof} Since the scalar curvature is non-positive it follows from the
Hamiltonian constraint that $k_{ab}k^{ab}\le (\tr k)^2$ and hence that
$\sum_{a}p_a^2\le 1$. The minimum of $\sum_{a}p_a^2$ subject to the constraint
$\sum_{a}p_a= 1$ occurs when all $p_a$ are equal, as follows from the method 
of Lagrange multipliers. Hence
\begin{equation}\label{k2min}
k_{ab}k^{ab}\ge \frac{(\tr k)^2}{n}.
\end{equation}
The minimum possible value of $p_a$ is $-\frac{n-2}{n}$, as has been proved
in \cite{rendall08}, p. 96. By symmetry this is also the minimum value of any 
other $p_a$. It can be concluded that the maximum eigenvalue 
$\lambda_{\rm max}$ satisfies 
\begin{equation}
\frac1{n}\tr k\le\lambda_{\rm max}\le -\frac{n-2}{n}\tr k.
\end{equation}
If the value of $\lambda_{\rm max}$ is fixed the inequality (\ref{k2min})
can be improved. Assume that $ \lambda_{\rm max}=\lambda_1$. Using
Lagrange multipliers again it can be seen that the minimum of 
$k_{ab}k^{ab}$ is attained when all $\lambda_a$ other than $\lambda_1$
are equal. Hence the lower bound for $k_{ab}k^{ab}$ is got by setting
\begin{equation}
\lambda_2=\ldots=\lambda_n=\frac{\tr k-\lambda_{\rm max}}{n-1}.
\end{equation}
It follows that $k_{ab}k^{ab}$ is bounded below by
\begin{eqnarray}
&&\lambda_{\rm max}^2+\frac{(\tr k-\lambda_{\rm max})^2}{n-1} \nonumber\\
&&=(\tr k)^2(np_{\rm min}^2-2p_{\rm min}+1)(n-1)^{-1}         \nonumber\\
&&=(\tr k)^2 P(p_{\rm min}).
\end{eqnarray}
The function $P$ is positive and decreasing on 
$\left[-\frac{n-2}{n},\frac{1}{n}\right]$. Let $p_0$ be the supremum in 
$t$ of $p_{\rm min}(t)$. Then by definition $p_{\rm min}(t)\le p_0$ for all
$t$. On the other hand the hypotheses of the theorem imply that 
$p_{\rm min}(t)\ge p_0-\Delta$ where $\Delta=p_0-\inf_{t\ge t_0}\min_a p_a(t)$. 
Assume that $p_0-\Delta<0$. (Geodesic completeness in the case $p_0\ge\Delta$ 
is treated in Theorem \ref{poscomplete}.) Now 
\begin{equation}
\partial_t (\tr k)=k_{ab}k^{ab}\ge P(p_0)(\tr k)^2.
\end{equation}
Integrating this in $t$ gives
\begin{equation}\label{trklower}
(\tr k)(t)\ge\frac{-P(p_0)^{-1}}{C+t}
\end{equation}
for a constant $C$ depending on $t_0$ and $\tr k(t_0)$. Let $q^a$ be the 
projection of the tangent vector to a causal geodesic onto the hypersurfaces
of constant $t$. Then
\begin{equation}
\frac{d}{dt}(g_{ab}q^aq^b)=2k_{ab}q^aq^b\le 2(p_0-\Delta)(\tr k)g_{ab}q^aq^b.
\end{equation} 
Replacing $\tr k$ in this inequality using (\ref{trklower}) and 
integrating gives
\begin{equation}
(g_{ab}q^aq^b)^{-\frac12}\ge C(1+t)^{-\lambda}
\end{equation} 
where $\lambda=\frac{\Delta-p_0}{P(p_0)}$. The affine parameter length of a 
causal geodesic up to time $\tau$ is given by
\begin{equation}
\int_{t_o}^\tau (\epsilon+g_{ab}q^aq^b)^{-\frac12}dt
\end{equation}
where $\epsilon$ is one for timelike geodesics parametrized by arc length and 
zero for null geodesics. This integral diverges provided $\lambda \le 1$. It
remains to show that this follows from the inequality
$\Delta\le\frac{9-n}{4n}$. Now
\begin{equation}
\frac{\Delta-p_0}{P(p_0)}\le 1
\end{equation}
is equivalent to
\begin{equation}
p_0^2+\frac{n-3}{n}p_0+\frac{1}{n}-\frac{n-1}{n}\Delta\ge 0.
\end{equation}
The expression on the left hand side of this inequality has its minimum at
$p_0=-\frac{n-3}{2n}$. Its value there is
\begin{equation}
\frac{n-1}{n}\left(\frac{9-n}{4n}-\Delta\right).
\end{equation}
This completes the proof of the theorem.

\vskip 10pt
For simplicity this theorem has been stated for the vacuum case only. 
In fact the theorem, and its proof, generalize straightforwardly to
the case where matter is present provided the matter satisfies the 
dominant and strong energy conditions and has reasonable evolution 
properties. The precise formulation of the last property is the matter 
continuation criterion (MCC) introduced in \cite{rendall08}. A variant
of Theorem \ref{gencomplete} is given by

\begin{theorem}\label{poscomplete}
Let $(M,g)$ be a locally spatially homogeneous vacuum spacetime of dimension 
$n+1$ whose symmetry is defined by a Lie group $G$ which admits no
left-invariant metrics of positive scalar curvature. If for some $t_0$
the generalized Kasner exponents satisfy $p_a\ge-\frac{1}{n}$ then the 
spacetime is geodesically complete.
\end{theorem}

\noindent
{\bf Proof} The proof proceeds in a similar way to that of the previous 
theorem. This time the inequality $\partial_t(\tr k)\ge\frac{(\tr k)^2}{n}$
implies that
\begin{equation}\label{trklower2}
(\tr k)(t)\ge -\frac{n}{C+t}
\end{equation}
and that
\begin{equation}
(g_{ab}q^aq^b)^{-\frac12}\ge C(1+t)^{-\lambda}.
\end{equation}
This completes the proof.

\vskip 10pt\noindent
This theorem also generalizes easily to the case with matter satisfying
the DEC, SEC and MCC. In the case $n=3$ the hypothesis on the $p_a$ is 
satisfied identically.
 
Next a large class of Lie groups will be exhibited for which 
the assumptions of Theorem \ref{gencomplete} are satisfied. They are 
obtained using products of three-dimensional Lie groups with Abelian 
groups of arbitrary dimensions.

\begin{theorem}\label{maintheorem}
Let $(M,g)$ be a spatially homogeneous solution of the vacuum Einstein 
equations in $n+1$ dimensions with $3\le n<9$. Suppose that the group 
defining the homogeneity is of the form $G_3\times (\R\rtimes \Z_2)^{n-3}$ 
for some three-dimensional Lie group $G$ whose Bianchi type is neither IX or 
VI${}_{-\frac19}$. If the solution is maximally extended to the future then
it is geodesically complete. 
\end{theorem}

\noindent
{\bf Proof} Under the hypotheses of this theorem Kaluza-Klein reduction can be 
used to relate the generalized Kasner exponents of the higher dimensional
vacuum model to the generalized Kasner exponents of the reduced 
Einstein-scalar field model in four dimensions. By Theorem \ref{mainscalar}
and the formulae (\ref{red1}) and (\ref{red2}) it follows that the 
generalized Kasner exponents of the $(n+1)$-dimensional spacetime converge 
as $t\to\infty$. This conclusion makes use of the fact that, since
$\frac{2\Omega}{3}=(\frac{\partial_{\tilde t}\phi}{\tr \tilde k})^2$, the 
convergence of $\frac{\partial_{\tilde t}\phi}{\tr \tilde k}$ is equivalent 
to that of $\Omega$. Hence $p_0=0$ and the hypotheses of Theorem 
\ref{gencomplete} are satisfied.

\section{Further examples}

In the previous section geodesic completeness was proved for models based
on Lie groups belonging to a certain class not admitting metrics of
positive scalar curvature. To go further it is necessary to enter into
the structure of more general Lie algebras. (A good general reference for
the theory of Lie groups and Lie algebras is \cite{varadarajan}.)
For this some definitions
are necessary. If $L$ is a Lie algebra denote by $[L,L]$ the subalgebra 
generated by the elements of $L$ of the form $[x,y]$ for $x$ and $y$ in $L$. 
Define $L^{(n)}$ recursively by $L^{(n+1)}=[L^{(n)},L^{(n)}]$ and $L^{(1)}=L$. 
If $L^{(n)}=\{0\}$ for some $n$ then $L$ is called solvable. The radical $R$ 
of a Lie algebra $L$ is an ideal of $L$ which is solvable and which contains 
all other solvable ideals. A Levi subalgebra is a subalgebra which, as a 
vector space, is complementary to the radical. It is semisimple which means 
that it has no non-trivial solvable ideals. A derivation of $L$ is a linear
mapping $D$ which satisfies $D([x,y])=[Dx,y]+[x,Dy]$ for all $x$ and $y$.
A particular type of derivations are the inner derivations which are of
the form $[x,\ ]$ for some element $x$ of the algebra.
Let $L_1$ and $L_2$ be two Lie algebras. Denote by ${\rm Der} L_2$ the 
vector space of all derivations of $L_2$. It is closed under commutators
and in this way acquires the structure of a Lie algebra. Let $\phi$ be a Lie 
algebra homomorphism from $L_1$ to ${\rm Der} L_2$. A Lie algebra $L$ can be 
defined as the direct sum of the underlying vector spaces of $L_1$ and $L_2$ 
with the Lie bracket defined by
\begin{equation}
[(x_i,y_1),(x_2,y_2)]=([x_1,x_2],[y_1,y_2]+\phi(x_1)(y_2)-\phi(x_2)(y_1)).
\end{equation}
The Lie algebra $L$ is called the semidirect sum of $L_1$ and $L_2$.
The Levi-Malcev theorem says that any finite-dimensional Lie algebra 
can be written as a semidirect sum of its radical $R$ with a semisimple Lie
algebra. 

Consider now the case of four-dimensional Lie algebras. It is known that
the only semisimple Lie algebras of dimension no greater than four are
$su(2)$ and $sl(2,\R)$. It follows that the only four-dimensional Lie
algebras which are not solvable are semidirect sums of the real numbers with
$su(2)$ and $sl(2,R)$. In fact the semidirect sum of the real numbers with a 
semisimple Lie algebra is isomorphic to a direct sum. To see
this note that every derivation of a semisimple Lie algebra is inner. Hence
the mapping $\phi$ defining the semidirect sum must be of the form
$\phi(x)=\phi_S(x)[y_0,\ ]$ for a linear map $\phi_S$ from $L_1$ to $\R$ 
and some $y_0\in L_2$. It then follows that the linear map $\psi$ from $L$ 
to itself defined by 
\begin{equation}
\psi(x,y)=(x,y+\phi_S(x)y_0)
\end{equation}  
is an isomorphism from the given Lie algebra to the direct sum. The group 
$SU(2)\times\R$ admits left-invariant metrics of positive scalar curvature 
because $SU(2)$ does. On the other hand it is well-known that the simply 
connected Lie group with Lie algebra $sl(2,\R)$ is diffeomorphic to $\R^3$. 
Hence the group corresponding to the Lie algebra $sl(2,\R)\oplus \R$ is 
diffeomorphic to $\R^4$. Any simply connected solvable Lie group is 
diffeomorphic to $\R^n$ for some $n$ and hence admits no metric of positive 
scalar curvature. Putting these facts together it can be concluded that the
only connected and simply connected four-dimensional Lie group which 
admits a metric of positive scalar curvature is $SU(2)\times\R$. This fact 
was noted in the thesis of Hervik \cite{hervik04}, who refers to a paper
of Patera et. al. \cite{patera}. In the latter a list of Lie algebras of 
dimensions 4 and 5 is given which are not direct sums and they are all stated
to be solvable. It may be noted in passing that on the basis of this the only 
types for which recollapse can occur in five space dimensions are 
$su(2)\oplus\R^2$ and $su(2)\oplus A_{2,1}$. A general criterion for the
existence of left-invariant metrics of positive scalar curvature in any
dimension has been given in \cite{hervik10}. The facts just mentioned 
suggest the following conjecture:

\vskip 10pt\noindent
{\bf Conjecture} Let $G$ be a connected and simply connected four-dimensional 
Lie group which is not isomorphic to  $SU(2)\times\R$. Then any spatially 
homogeneous solution of the vacuum Einstein equations with this symmetry 
group which is expanding at some time is future geodesically complete.    

Theorem \ref{maintheorem} proves this conjecture for Lie algebras which are
the direct sum of a three-dimensional Lie algebra and the real numbers
and metrics which have an additional discrete symmetry. Another case where 
something is known is that where the four-dimensional algebra is a direct sum 
of two non-Abelian subalgebras \cite{andersson07}. In the notation of 
\cite{patera} and \cite{hervik02} this is the case $A_{2,1}\oplus A_{2,1}$. The 
metrics covered
by the theorem of \cite{andersson07} are not the most general left-invariant
metrics. They have a six-dimensional symmetry group. By Theorem 3.1 of
\cite{andersson07} the quantities $P$ and $Q$ converge to positive limits
as $t\to\infty$ and these are essentially the generalized Kasner exponents. 
Hence Theorem \ref{poscomplete} above applies. Similar results are obtained
in higher dimensions. These apply for example to Lie algebras of the form 
$A_{2,1}\oplus\R^n$ for any $n$. The limits of the generalized Kasner 
exponents are non-negative. It is once again the case that only metrics
with additional symmetry are included.   

The Lie algebras which define Bianchi models of class A are those which are
unimodular, i.e. those for which the trace of the structure constants is
zero. A four-dimensional Lie algebra which is the direct sum of a 
three-dimensional algebra with the real numbers is unimodular if and only
if the three-dimensional Lie algebra is of class A. The unimodular 
four-dimensional Lie algebras which are not of this form can be read off
from Table 1 in \cite{hervik02}. In the notation used there they are
$A_{4,1}$, $A^{-2}_{4,2}$, $A^{pq}_{4,5}$ for $p+q=-1$, $A^{pq}_{4,6}$ for
$p+2q=0$, $A_{4,8}$ and $A_{4,10}$. Some explicit solutions are known. In
the cases $A^{-2}_{4,2}$ and $A^{pq}_{4,5}$ explicit diagonal solutions are
given in \cite{demaret85}. In both cases the generalized Kasner exponents
converge to $(1,0,0,0)$ and thus these solutions are future geodesically
complete by Theorem \ref{poscomplete}. In the case $A_{4,8}$ a solution which 
is diagonal and self-similar is given in \cite{christodoulakis}. Its
generalized Kasner exponents are time-independent. They are equal to 
$(\frac34,-\frac14,\frac14,\frac14)$. Theorem \ref{poscomplete} applies to
this solution and it is interesting to note that this is a borderline case 
for that theorem. In the case $A_{4,10}$
an explicit solution is mentioned in \cite{christodoulakis} but the authors
note that it has a higher dimensional symmetry group and can also be 
considered as corresponding to the Lie algebra which is the direct sum of
the Bianchi type II Lie algebra with the real numbers. It admits a reflection
symmetry of the type which makes it belong to the class covered by Theorem 
\ref{maintheorem}.

There is a way of fitting many of these Lie algebras in a common geometrical
framework. In \cite{rendall95a} the notion of local $U(1)\times U(1)$ 
symmetry was introduced and it was shown how the Bianchi types I, II,
VI${}_0$ and VII${}_0$ can be identified among spacetimes with two
commuting spacelike local Killing vectors and a global topology which is
a bundle over the circle with fibre $T^2$. The different Bianchi types are
distinguished by the algebraic structure of a certain $2\times 2$ matrix
with unit determinant. If both eigenvalues are real with unit modulus and the 
matrix is diagonalizable then type I is obtained. A matrix which is not 
diagonalizable gives rise to type II. When the eigenvalues are real and 
distinct type VI${}_0$ results. Finally type VII${}_0$ is obtained when the 
eigenvalues are imaginary. The same analysis can be repeated for one 
dimension higher, although it does not cover all relevant algebras. In this 
case local $U(1)\times U(1)\times U(1)$ symmetry is the basic input and 
different cases are classified by the algebraic properties of a $3\times 3$ 
matrix with unit determinant. If there is a real eigenvalue with unit modulus 
and the corresponding Jordan block is one by one then a case is obtained 
where the Lie algebra is the direct sum of one of the algebras seen in the 
three-dimensional case with the real numbers. The only other case where all 
eigenvalues have are real with unit modulus is that of a $3\times 3$ Jordan 
block. This 
gives the algebra $A_{4,1}$. When the eigenvalues are real and all distinct 
type $A^{pq}_{4,5}$ results. In the case that there is pair of non-real 
eigenvlaues the Lie algebra which arises is $A^{pq}_{4,6}$. If there is a 
non-trivial Jordan block corresponding to a double eigenvalue then 
$A^{-2}_{4,2}$ occurs\footnote{Remarks related to the material of
this paragraph are contained in an unpublished manuscript of S. Hervik}.
This construction does not apply to types $A_{4,8}$ and $A_{4,10}$ since
it can be shown straightforwardly that neither of these has a 
three-dimensional Abelian subalgebra. This point of view shows that in all 
the cases where the Lie algebra has a three-dimensional Abelian subalgebra 
the group can be made to act on the universal cover of a spacetime with a 
compact Cauchy surface by isometries of the  pull-back of the spacetime 
metric. The topology of the Cauchy surface is that of a bundle over $S^1$ 
with fibre $T^3$.

A convenient feature of vacuum Bianchi models of class A is that initial 
data whose components in a suitable basis are diagonal lead to solutions 
which are also diagonal in that basis. Call a basis of this kind a 
canonical basis. It is moreover the case that for any initial data of 
Bianchi class A there is a canonical basis in which it is diagonal.
These facts together mean that in order to understand the dynamics of 
vacuum models of Bianchi class A it is enough to understand the dynamics
of diagonal models. The fact that models which are initially diagonal 
remain diagonal can be proved using automorphisms of the Lie algebras
involved. For convenience let the linear transformation which multiplies 
$e_i$ by $(-1)^{a_i}$, $a_i\in\{0,1\}$, be denoted by $(a_1a_2a_3)$. In 
a canonical basis the transformations $(110)$, $(101)$ and $(011)$ are
automorphisms. In other words, they leave the structure constants 
invariant. Data which are diagonal in the given basis are invariant 
under these transformations. Because the structure constants are
invariant the Einstein evolution equations are also invariant under
these transformations. It follows from the unique determination of
solutions by data that invariant data lead to invariant solutions.
A solution which is invariant under all these transformations is diagonal.  
It would be nice if this argument could be generalized to the case of
unimodular Lie algebras in higher dimensions. This is no problem if the
Lie algebra is a direct sum of a three-dimensional Lie algebra with the 
real numbers. For in that case the automorphisms of the three-dimensional
case can be trivially extended to get (in a hopefully obvious notation)
the automorphisms $(1100)$, $(1010)$ and $(0110)$. For the other Lie
algebras this method has very limited success, at least if the bases in 
which the Lie algebras are presented in \cite{hervik02} are taken as 
candidates for a canonical basis. The only case in which there are enough 
automorphisms obtained by changing the signs of some of the basis vectors 
to make the argument go through is $A^{pq}_{4,5}$. For that Lie algebra
any automorphism must fix $e_4$ but the signs of the other basis vectors
can be changed independently. In other words it is possible to use the
automorphisms $(1000)$, $(0100)$ and $(0010)$. The Wainwright-Hsu system
has been very useful for the study of Bianchi class A models. Its 
simplicity relies on the diagonalization property and so it may be that 
nothing of comparable power is available for general unimodular Lie
algebras in any higher dimension.

The theme of the present paper is the late-time behaviour of cosmological
models but the behaviour near the initial singularity is also of great
interest. For the latter subject the issue of diagonalization plays a
very important role. In spacetime dimension four it is believed that generic 
spatially homogeneous solutions of the Einstein vacuum equations show
oscillatory (Mixmaster) behaviour near the initial singularity and this
has been proved rigorously in some cases. (For Bianchi type IX see
\cite{ringstrom01a}.) In $4+1$ dimensions it seems that
this is still true but that restricting to the special case of diagonal
models removes the generic oscillations \cite{demaret89}. In particular,
generic oscillations can occur for the group $SU(2)\times \R$ due to the 
influence of the non-diagonal degrees of freedom. The situation is similar
in dimensions up to $9+1$ but beyond that the oscillations are no longer
generic. For some rigorous results on this see \cite{damour02}.

\section{Conclusions and outlook}

In the preceding sections it has been proved that spatially homogeneous
solutions of the Einstein equations in various dimensions with symmetries
defined by a large class of Lie groups are geodesically complete in the 
future. The case considered in most detail is that of spacetime dimension
five. Even there it is to be hoped that something much more general can be
proved. A suggestion for a generalization of this kind is formulated as a
conjecture in Section 5. Making a corresponding conjecture in general 
dimensions is dependent on a sufficently good understanding of the 
structure of general Lie algebras. A guide in formulating the conjecture
was to restrict to those Lie groups which do not admit a metric of 
positive scalar curvature. Another direction of possible generalization 
is to consider spatially homogeneous spacetimes where the Lie group 
defining the symmetry is necessarily of a dimension higher than the space
dimension. These generalize the Kantowski-Sachs models in spacetime 
dimension four. In five spacetime dimensions the list of possibilities
is as follows \cite{hervik02}: the four-sphere $S^4$ with symmetry group
$SO(4)$, the complex projective plane $\C\P^2$ with an eight-dimensional
symmetry group, $S^2\times S^2$ with a six-dimensional symmetry group,
$S^2\times\R^2$ and $S^2\times H^2$ with five-dimensional symmetry groups.
All of these examples admit an invariant metric with positive scalar 
curvature and so are not promising candidates for allowing spacetimes 
which are future geodesically complete. The theorem of
B\'erard-Bergery \cite{berardbergery} mentioned at the end of Section 2
extends to homogeneous spaces which are more general than Lie groups
and may be helpful in investigating generalizations of Kantowski-Sachs
models in higher dimensions.   

In the previous sections the cosmological constant $\Lambda$ has always been 
assumed to vanish. It is likely that the theorem of Wald \cite{wald} on the 
late-time asymptotics of spatially homogeneous cosmological models with
$\Lambda>0$ extends to arbitrary dimensions with the essential assumption
(in the vacuum case) being the absence of metrics of positive scalar
curvature. Combining this with the ideas in \cite{lee} should give 
geodesic completeness. These ideas have, however, not yet been worked out
in detail.

There are very few results on future geodesic completeness for 
inhomogeneous spacetimes which do not rely on choosing initial data which
are close to those of a known solution (small data results). This applies even 
in the a priori simpler case $\Lambda>0$. In fact the only result of this 
kind for vacuum spacetimes is that of Ringstr\"om \cite{ringstrom04}. It could 
be that obtaining a better understanding of geodesic completeness for 
homogeneous models in higher dimensions would provide new ideas for 
understanding the inhomogeneous case in spacetime dimension four.

It is interesting to ask what happens when there is a metric of positive
scalar curvature. Is there always recollapse in that case? It seems that
very little is known about this in spacetime dimensions greater than
four. There are also few results available for inhomogeneous solutions in four 
dimensions. Perhaps a better understanding of the homogeneous case in 
higher dimensions could help to change this. In particular, higher 
dimensions could be a good place to look for a counterexample. 

The long-time behaviour of solutions of the vacuum Einstein equations in 
higher dimensions gives rise to a rich variety of problems concerning
the dynamics of the gravitational field. In this paper we have obtained
theorems on some of the simpler of these, while pointing out promising 
directions for future research. We have also emphasized that one motivation
for studying these questions is the insights they may bring for the 
Einstein-matter equations in four dimensions.

\end{document}